# Acceleration of the condensational growth of water droplets in an external electric field

**Dmitrii N. Gabyshev[1*], Alexander A. Fedorets[1**], Nurken E. Aktaev[1***], Otto Klemm[2****], Stepan N. Andreev[3*****]**

[1] University of Tyumen, Volodarsky Str. 6, 625003 Tyumen, Russia
[2] Westfälische Wilhelms-Universität Münster, Climatology Research Group, Heisenbergstr. 2, 48149 Münster, Germany
[3] Federal Center of Expertize and Analyzis, ul. Talalikhina 33 bd. 4, 109316 Moscow, Russia

[*] E-mail: d.n.gabyshev@utmn.ru, gabyshev-dmitrij@rambler.ru (corresponding author)
[**] E-mail: fedorets@utmn.ru
[***] E-mail: n.e.aktaev@utmn.ru
[****] E-mail: otto.klemm@uni-muenster.de
[*****] E-mail: andreev-sn@fgbnuac.ru

The condensational growth of spherical water microdroplets is studied in a laboratory setup and with a mathematical model. In the experiment, droplet clusters are kept in a freely levitated state within an upward-oriented flow of water vapor. In the presence of an electrostatic field of $1.5 \cdot 10^5$ V m$^{-1}$, droplet growth is accelerated by factors 1.5 to 2.0 as compared to conditions without any external electric field. Presumably water molecules in the ambient air are accelerated through the presence of the electric field. A kinetic model to predict the acceleration of condensational growth confirms this hypothesis to be feasible. The droplets themselves are polarized so that the deposition of steam molecules is facilitated in the electric field. The simplifications and limitations of the model are discussed.

**Highlights**

* Electric field intensifies the condensational growth of levitated droplets
* Condensational growth depends linearly on the strength of an electric field
* Droplet polarization plays an important role in electro-condensation
* Condensational growth in electric field occurs anisotropically
* The droplet cluster technology is an effective tool for studying droplet condensation

**1. Introduction**

The condensation process of water in an electric field has attracted increasing attention over the last decades [1–9]. Strong electric fields do exist within thunderstorm clouds [10, 11]. Taylor derived an exact theory of the deformation and disintegration of drops in strong electric fields [12] and thus set the basis for further studies in this area [13–16]. In cumulus clouds, the radii of droplets range from a few μm to over 100 μm, with a typical range of 5 – 20 μm. In the lower regions of the clouds, the growth of particles occurs mainly due to condensation. Coagulation is typical for the central zone of a cloud, typically reaching up to half the vertical extent of a cloud [17]. There are a number of experimental facilities to mimic microphysical processes occurring in clouds and fogs, including condensation [18–21].

The sheer fact of the intensification of condensation by an external electric field is known and used in some practical applications [22–24]. The process is described not only to water but also for other liquids [25–27]. Nevertheless, it is still a challenge to observe the condensational growth of individual droplets on real-time scales, in the real world or even under laboratory conditions. Even more, the quantification of the enhancement of the condensational droplet growth in the presence of an external electrical field such as in a thunderstorm cloud is an unresolved issue as of now.

The scope of this study is to develop further understanding of the condensation process of water droplets in electric fields with field strengths similar to those of atmospheric thunderclouds. We employ the laboratory technique of levitated droplet clusters [28] in combination with a kinetic model. Qualitative agreement of experimental and modeling results will indicate that the basic processes are understood and that the enhancement of the condensation process in clouds through electrical fields may be strong.

Levitated droplet clusters are array-type aerosols [29], which can be used for retaining microdroplets in a levitated state over a locally heated surface of a liquid. It is the pressure of the flow of evaporating liquid that keeps the droplets in a stationary, levitated state. The aqueous microdroplets form spontaneously in laboratory conditions, yet are not injected or inserted. They are formed through spontaneous heterogeneous condensation. Further, slow condensational growth occurs through their life time, eventually ensuring a rather extended lifetime in the order of tens of seconds up to minutes. In this study, the established technique is further developed towards the study of the influence of an external electric field on the condensation process.

At the outset of observations of levitated clusters, it was found that the droplet growth follows closely Maxwell's diameter-square law [30, 31], meaning that the droplet area grows linearly with time. Later, a technology was developed to suppress condensational growth through the exposure to infrared radiation [32, 33]. The dependence of the growth rate of droplets on their number in the cluster was studied in [34]. More recent research focused on the influence of periodic heating of the subjacent water layer on the droplets' growth [35]. These were the starting points to study the droplets' growth under the influence of an external electric field [36].

In the present study, the condensational growth of droplets in an electric field is described with a simple kinetic model. The model is based on the polarization of steam molecules around a single microdroplet, which itself is polarized in the external field. Assumptions and potential reasoning of disagreement between experimental and model data are presented.

## 2. Details of experiment
### 2.1. Experimental setup

In the general experimental setup, a stable, levitated droplet cluster forms spontaneously over a locally heated layer of distilled water (Fig. 1a). The water layer is on a sitall substrate of 400 μm thickness, which's bottom side is heated by use of a laser beam, which is focused on a 1-mm-diameter spot. The beam source is a KLM-H808-600-5 semiconductor laser (FTI-Optronic, Russia). The beam power is monitored by a power and energy meter console PM200 equipped with a S401c sensor (Thorlabs, USA). The thickness of the water layer was maintained at 400 ± 2 μm and the water was thermostated at 18.0 ± 0.5 °C. Under these conditions, there is a constant flux of evaporating water from its upper surface upward. Droplets appear spontaneously due to condensation of steam, with aerosol particles in the laboratory air acting as condensation nuclei. Individual droplets form a hexagonal cluster, which levitates at a distance of about one droplet diameter above the water surface (Fig. 1b). Droplet and cluster lifetimes are typically a few tens of seconds. Video recordings of the cluster are made using an AXIO Zoom.V16 stereo microscope (Zeiss, Germany) equipped with a PCO.EDGE 5.5C speed camera (PCO, Germany). The technique of a cluster's images analysis is described in detail in [37]. The thermal imaging camera Flir A655sc (Flir, USA) allows to control the temperature of the surface of the water layer. Further details of the experimental setup are described in [28, 36–39].

For the establishment of an external electrical field, electrodes are positioned below and above the sitall and droplet arrangement. The configuration and position of the electrodes are shown in Fig. 1a. The bottom electrode is a metal washer (outer diameter 8.9 mm, diameter of the central hole 4.5 mm, thickness 0.7 mm) located just below the sitall substrate. The upper electrode is a disk (outer diameter 78 mm, diameter of the hole 6.5 mm) made of metallized (copper layer, 18 μm thick) fiberglass.

The distance $l_U$ between the electrodes in all experiments remained fixed at 8 mm. A positive or negative potential can be applied to the upper electrode. An external electrostatic field is generated by a high-voltage source HVLAB3000 (ET Enterprises, UK). The applied voltage $U$ was between 0.200 kV and 3.000 kV with a step increment of 1 V. An electric field between zero and about $5 \cdot 10^5$ V m$^{-1}$ can thus be established (see Appendix). The polarity was typically chosen to be with a positive potential on the upper electrode (see Fig. 1, $U^+$). For some experiments, the polarity was reversed ($U^-$ at the upper electrode). In these cases, the term "polarity reversal" will be used. Fig. 1a also shows the cluster's geometry parameters $D$ (droplet diameter), $L$ (distance between the centers of neighboring droplets) and $h$ (distance from the water surface to the droplet's base).

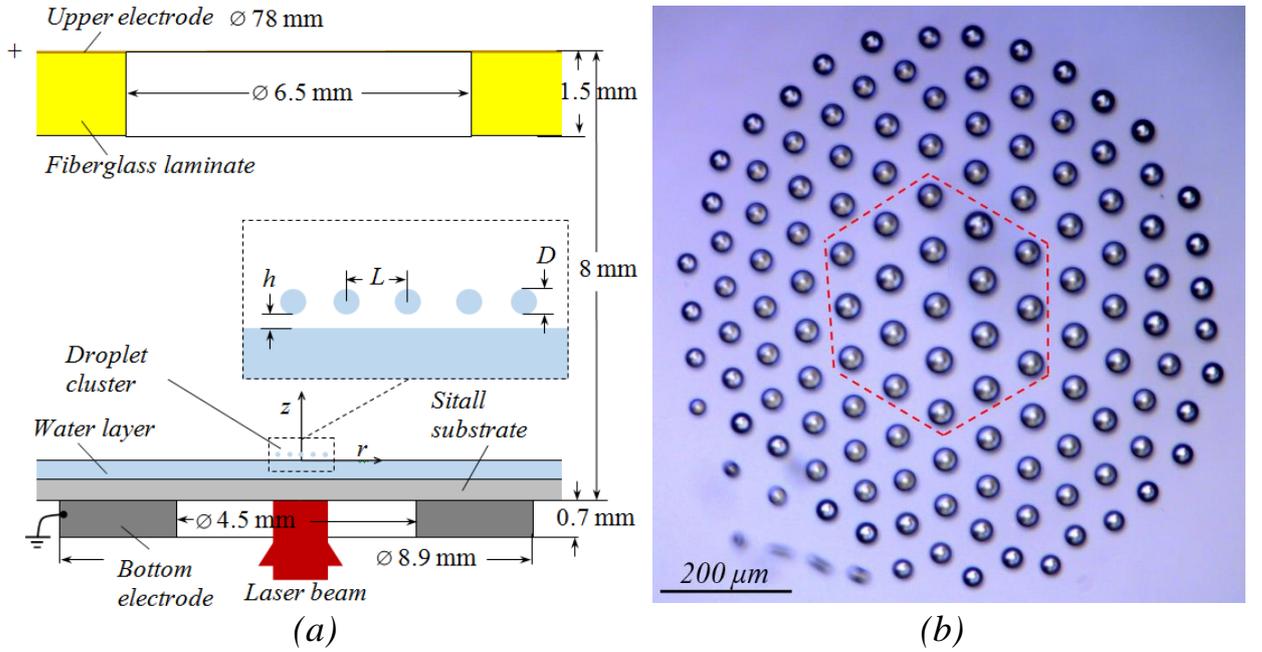

**Fig. 1.** *(a) Scheme of the experimental facility and electrodes' configuration. The zoomed-in region shows the droplet cluster from side with its geometric parameters; (b) Droplet cluster, top view ($U^+=400$ V). The dashed line frames a group of 19 droplets in the cluster center, of which the average condensational growth rate of the droplets was measured.*

## 2.2. Application of the electric field

Each droplet of a cluster participates in both evaporation and condensation [28, 40]. Either of these processes prevails at the bottom droplets' hemispheres, while the other process prevails at the top hemispheres. Overall, condensation prevails and the droplets grow continuously. In the absence of an external electric field, the surface area of a cluster's droplet increases linearly with time [31]. The condensational growth rate is quantified as the increase of droplet surface area with time.

The droplets remain spherical as a significant deformation does not occur before field strengths of about $\sqrt{\sigma/\varepsilon\varepsilon_0 D}$ ~ $10^6$ V m$^{-1}$ are reached ($\sigma$ is the surface tension, $\varepsilon_0$ is the electric constant, $\varepsilon$ is the dielectric constant of droplets). However, in an external electric field, the size of droplets capable of stable levitation is significantly reduced due to electrokinetic forces which tend to bring the cluster out of mechanical equilibrium [41–42]. The lifetime of droplets is reduced under such conditions. When $U > 700$ V, the remaining time window does not allow the formation of stable clusters with a large number of droplets. Therefore, the upper value of the supplied voltage was limited to 700 V (corresponding to $E = 1.53 \cdot 10^5$ V m$^{-1}$). The experimental data on the condensational growth rate were averaged over the group of droplets in the cluster center (Fig. 1b).

## 3. Model description

In this section, we will estimate an addition to the condensational rate $\dot{S}_0$ when the electric field is turned on. We will consider the process of condensational growth of a single droplet to be of quasi-stationary nature, when all the thermodynamic parameters of the medium and the droplets in it remain approximately constant. We follow the molecular-kinetic scheme proposed in [43] with some additions as shown below. The charge $q$ of a droplet of radius $R$ is determined by:

$$q = -k R^2,$$

where $k$ is a coefficient of electrification depending on environmental conditions such as the cloud shape and stage of its development [44]. It can be assumed to be equal to $3.364 \cdot 10^{-7}$ C m$^{-2}$ or 1 e.s.u. cm$^{-2}$ (CGSE). The sign of the droplet's own charge is defined to be negative in accordance with the last data on the droplet clusters [36].

Let the external electric field $\mathbf{E_0}$, in which the droplet is located, be homogeneous, as deduced in Appendix. The field near the surface of the droplet is equal to the field around a charged sphere with a dielectric constant $\varepsilon$ [45]. The radial components of the field are equal:

$$\mathbf{E_r} = \mathbf{j}\left(1 + 2\frac{\varepsilon - 1}{\varepsilon + 2}\frac{R^3}{r^3}\right)E_0 \cos \Theta \ , \ \mathbf{E_q} = \mathbf{j}\frac{1}{4\pi\varepsilon_0}\frac{q}{r^2} , \tag{1}$$

with $E_0 > 0$ and the components $\mathbf{E_r}$ and $\mathbf{E_q}$ being directed along the normal $\mathbf{j}$ to the surface of the spherical droplet, and $\Theta$ being the angle between the normal and the external field $\mathbf{E_0}$. The polarization of the droplet depends simultaneously on both the external field and the field created by the charges of the droplet itself [46]. Charges, which are distributed uniformly over the surface of the sphere, will not create field strengths within the sphere. However, a field can be created by charges distributed over the volume of the droplet or concentrated on the condensation nucleus. For a droplet with a diameter of about 50 μm, the charge is about $10^2 – 10^3$ units of the elementary charge [40, 44, 47, 48]. Therefore, the field $\mathbf{E_q}$ is of the order of $10^3$ V m$^{-1}$ on the surface of the droplet. This is by orders of magnitude smaller than the external field strength $\mathbf{E_0}$ in our experiments or in thunderstorm clouds ($10^4 – 10^6$ V · m$^{-1}$ [48]). Numerical calculations in the PiC software package KARAT [49] confirm that noticeable deviations from equation (1) are observed only when the charge of droplets is 2 orders of magnitude larger than that established in real experiments (the charge in these calculations is considered to be distributed uniformly over the droplet volume). It is therefore an appropriate simplification to neglect the self-polarization of the droplet associated with the field $\mathbf{E_q}$, and to assume that the polarization is due only to the external field $\mathbf{E_0}$.

The water vapor molecules are dipoles with a constant dipole moment $p_0 = 6.188 \cdot 10^{-30}$ C · m, which is virtually constant at field strengths of up to $10^8$ V m$^{-1}$ [50]. However, near the surface of a droplet, its orientation changes vigorously. According to the model of a rigid rotor [51], the dipole moment $p_0$ starts to oscillate with a frequency of 2.8 · 10¹⁰ Hz when entering into a field of about 2 · 10⁵ V m⁻¹. The molecule has enough time to make several oscillations as the frequency of molecular collisions is about 1 · 10¹⁰ s⁻¹ and thus less than the oscillation frequency. The average value of the projection $\langle p_0 \rangle \approx p_0^2 E/3kT$ onto the electric field direction is thus negligible at room temperature for most water molecules [52]. Therefore, in the model, only the induced dipole moment plays a role [43]:

$$\mathbf{p_i} = 4\pi\varepsilon_0 a \mathbf{E},$$

where $a = 5 \cdot 10^{-29}$ m³ is the electronic polarizability of a water molecule.

The potential energy of the dipole in the electric field $\mathbf{E} = \mathbf{E_r} + \mathbf{E_q}$ is equal to:

$$\Pi = -\mathbf{p_i} \cdot \mathbf{E} = -4\pi\varepsilon_0 a \mathbf{E}^2.$$

The increment of the kinetic energy of the water molecule at the mean free path $\langle \lambda \rangle$ in the near-surface layer of the atmosphere around the droplet is equal to the loss of potential energy. The initial velocity of a water molecule is set to zero immediately after the last intermolecular collision [43]. This assumption is valid if the kinetic energy of the translational motion during an intermolecular collision is entirely converted to other types of energy [53]. However, not every water vapor molecule moves for the full length $\langle \lambda \rangle$ after each collision. The last collision of a molecule may have occurred either at the distance $\langle \lambda \rangle$ or even in closer proximity to the droplet surface. In order to take the likelihood of the distances of the last collision from the water surface into account, the velocity of a typical molecule should be computed with half of the length $\langle \lambda \rangle$. Further,

$$\Pi(R, \Theta) - \Pi\left(R + \frac{\langle \lambda \rangle}{2}, \Theta\right) = \frac{m_0 v_\perp^2}{2}, \qquad (2)$$

where $m_0$ is mass of a water molecule, $v_\perp$ is the normal component of the molecular velocity, in the calculation of which only the loss of potential energy along the normal is taken into account:

$$\Pi = -4\pi\varepsilon_0 a \mathbf{E}_\perp^2 = -4\pi\varepsilon_0 a \left|\mathbf{E_r} + \mathbf{E_q}\right|^2.$$

The additional mass flow of vapor d$m$ precipitated on a droplet during the time d$t$ and caused by the drift of molecules in an electric field is equal to [43]:

$$\frac{dm}{dt} = \int_S \rho_v v_\perp \, dS = \rho_v \int_0^\pi v_\perp(R,\Theta) 2\pi R^2 \sin\Theta \, d\Theta,$$

where $\rho_v$ is the absolute humidity of the air, d$S$ is the surface area element, $S$ is the total surface area of the droplet. The area growth rate is equal to:

$$\frac{dS}{dt} = \frac{\rho_v}{\rho_w} \frac{2}{R} \int_0^\pi v_\perp(R,\Theta) 2\pi R^2 \sin\Theta \, d\Theta = 4\pi R \frac{\rho_v}{\rho_w} \int_0^\pi v_\perp(R,\Theta) \sin\Theta \, d\Theta.$$

The radius growth rate is equal to

$$\frac{dR}{dt} = \frac{1}{4\pi R^2} \frac{\rho_v}{\rho_w} \int_S v_\perp(R,\Theta) 2\pi R^2 \sin\Theta \, d\Theta = \frac{1}{2} \frac{\rho_v}{\rho_w} \int_S v_\perp(R,\Theta) \sin\Theta \, d\Theta.$$

The last expressions are implicit, because $R$ is both on the left and on the right hand side of equation 6. After rearrangement, the time is expressed as:

$$t = \frac{\rho_w}{\rho_v} \int_0^R \frac{2}{\int_0^\pi v_\perp(R',\Theta) \sin\Theta \, d\Theta} \, dR'. \qquad (3)$$

Generally speaking, equation (3) can be obtained in a more accurate form by taking into account the initial diffusion mass flow $\dot{S}_0$ in the absence of the field:

$$t = \int_0^R \left[ \frac{\dot{S}_0}{8\pi R'} + \frac{1}{2}\frac{\rho_v}{\rho_w} \int_0^\pi v_\perp(R',\Theta) \sin\Theta \, d\Theta \right]^{-1} dR'. \qquad (4)$$

Equation (4) describes the dependence of $t$ on $R$ or on $S$ (replacing the upper limit of the integral as $R = \sqrt{S/4\pi}$).

Further, neighboring droplets within the levitated array of droplets exert an influence on each other and can thus not be considered as being independent from each other. In earlier experiments [34], the growth rate of a droplet surrounded by other droplets was found to be reduced by almost $\gamma = \frac{1}{2}$ as compared to single droplets. This effect is caused by the competition of neighboring droplets for the water vapor and by an increased aerodynamic resistance of the cluster array. These effects need to be taken into account when calculating the final growth rate of droplets $\dot{S}$ below, introducing the empirical factor $\gamma = \frac{1}{2}$ before the second term within square brackets in (4) in addition to the numerical factor $\frac{1}{2}$ already existing there. More precisely, the velocity $v_\perp$ under the integral should be multiplied by the factor $\gamma$. Note that this approach does not apply to droplets at the cluster's rim, because the conditions for condensation are not homogeneous in this region. Therefore, our experimental data were obtained and averaged only on the central group of droplets (Fig. 1b).

## 4. Results and Discussion

For the computation of the surface area as a function of time (4), an air temperature $T$ of 60 °C (corresponding to the water layer temperature measured directly under the droplet cluster), a droplet temperature 58 °C, and a water dielectric constant $\varepsilon = 67.39$ at this temperature, an absolute humidity 0.1306 kg·m$^{-3}$, a mean free path in air $6.443 \cdot 10^{-8}$ m, as well as the water density 984.2 kg·m$^{-3}$ were used. The equations were resolved numerically. The results are shown in Fig. 2–4.

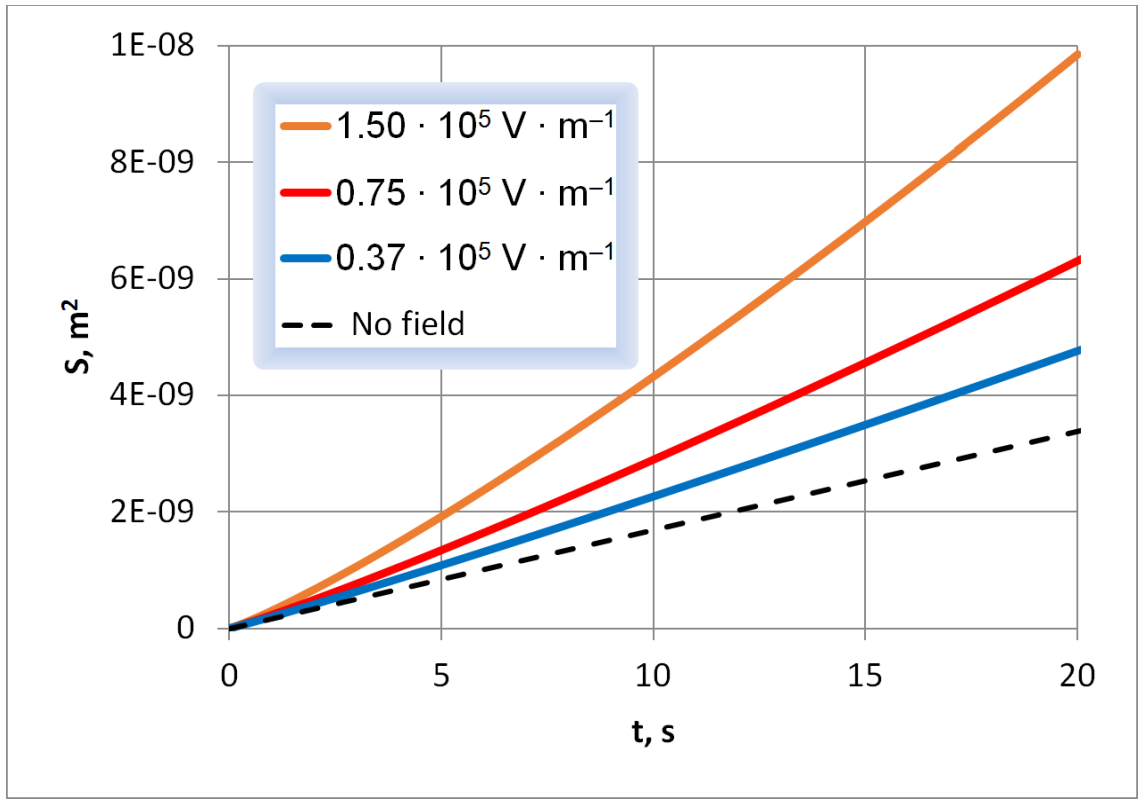

**Fig. 2.** *Surface area of a droplet S as function of time t according to (4).*

Since Fig. 2 shows that the growth rate gradually changes with time (the larger the droplet, the stronger the polarization field around it), Fig. 3 contains the average growth rate for a droplet with a radius of 15 μm (final area S divided by the total growth time *t*, when $R = 15$ μm). This radius is enough for the condensation growth to be replaced by the gravitational coagulation in natural clouds.

The results of the experiments, in which the rates of condensational growth of clusters' droplets in an electric field were measured, are summarized in Fig. 3. The experimental baseline is the measured condensational growth rate in the absence of an electric field is $\dot{S}_0 = 169.62 \cdot 10^{-12}$ m² s⁻¹.

The electric field significantly affects the growth rate of droplets. Droplets grow faster in the presence of electric fields, and the growth rates increase with increasing field strength. The polarity of the electrodes also plays a role [36]. For example, at a voltage of $U^- = 0.7$ kV (corresponds to $E = 1.53 \cdot 10^5$ V m⁻¹), the growth rate is $\dot{S}^- = 1.91\,\dot{S}_0$, while in the case of $U^+ = 0.7$ kV, the rate is only $\dot{S}^+ = 1.56\,\dot{S}_0$.

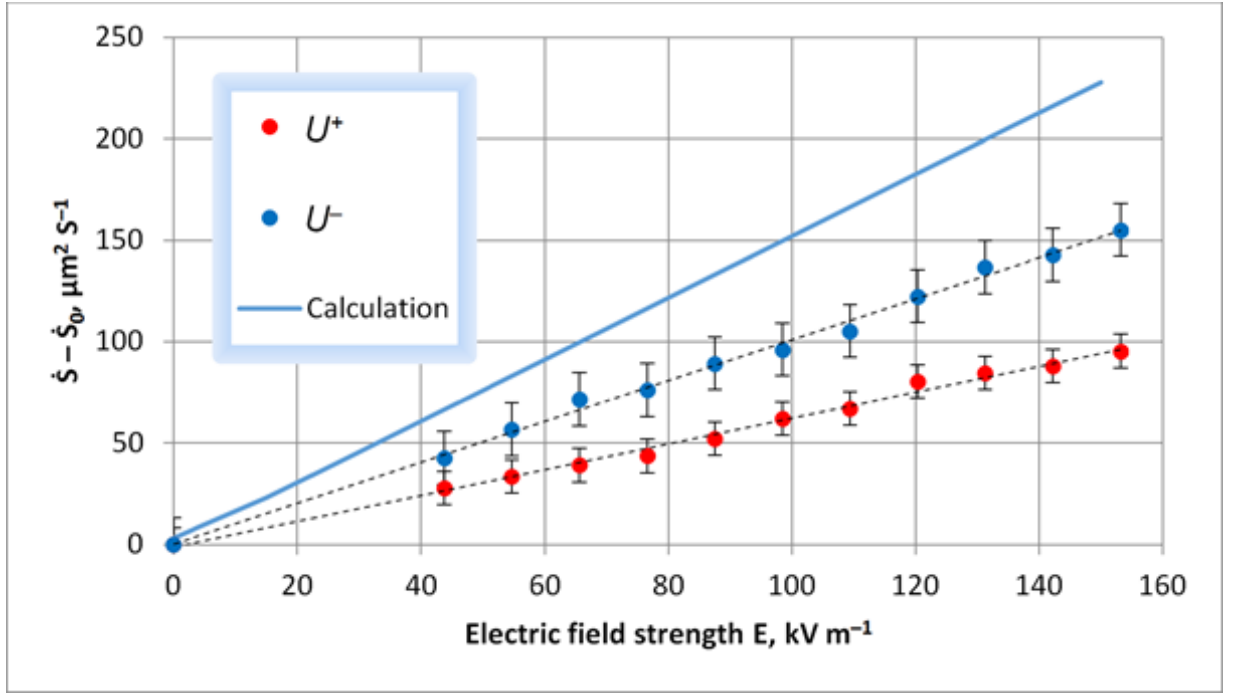

**Fig. 3.** *Increase of the average growth rate ($\dot{S} - \dot{S}_0$) as a function of the field strength E for a droplet of 15 μm radius.*

Fig. 3 demonstrates a linear dependence of the difference $\dot{S} - \dot{S}_0$ on the strength *E*. The blue line represents the calculation. There is in qualitative agreement with the measured data (blue and red dots in Fig. 3). The results of both the computation and the measurements show that the droplet will reach the same radius in less time in the stronger field. However, the modeled growth differences $\dot{S} - \dot{S}_0$ are by factors of 1.5 and 2.5 larger than those measured with upward ($U^-$) and downward ($U^+$) oriented fields, respectively. The incline of the calculated graph is almost independent on the coefficient of electrification *k* due to the weak contribution of the field of the own droplet's charges to the drift motion of precipitated molecules. The main contribution is due to the external electric field and the droplet's polarization.

There are several possible sources of the model inaccuracy. The most significant simplification of the model was the assumption of immobility of a droplet in the environment. Meanwhile, a vertically motionless droplet in a droplet cluster is embraced by the steam-air mixture, which is equivalent to the translational motion of the droplet in a fixed medium. A typical velocity of the flow supporting the droplet cluster is about 0.1 m s$^{-1}$ [39]. It exceeds the maximal drift velocity 0.0123 m s$^{-1}$ of the molecules falling onto the droplet (Fig. 4). Presumably, the steam-air flow just blows a significant part of electrically driven molecules upwards. Any solution for this type of phase transition requires a simulation of the steam-air flow around the droplets with a higher spatial resolution. This non-trivial task is beyond the scope of this contribution [54–60] and it requires further research.

Note that a strong electrostatic field should also affect the evaporation rate of water [7, 61, 62] and thus the properties of the steam-air flow itself, in which the cluster levitates. It has, for example, been shown that an external field of 10$^9$ V m$^{-1}$, oriented perpendicular to a water surface, enhances the vapor flow approximately by a factor of two [7]. Obviously, the intensity of this effect scales linearly with the field strength. Therefore, the effect should be proportionally (by a factor of $10^4$) less in our experiments, in which the field strength was in the order of $10^5$ V m$^{-1}$. Therefore, it is appropriate to neglect the influence of the electric field on the evaporation of a layer of water under the droplet cluster.

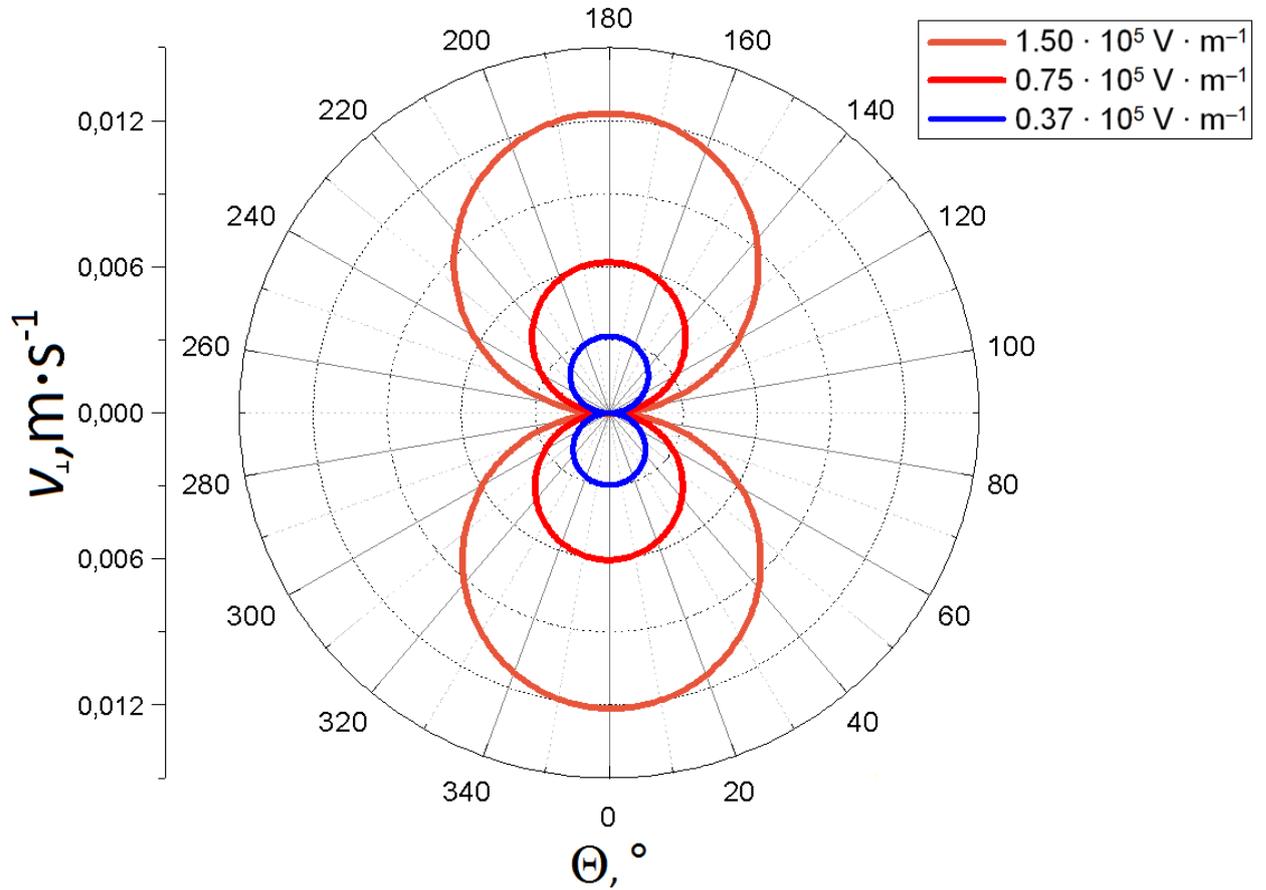

*Fig. 4. Angular distribution of molecular drift deposition velocities on a droplet (R = 15 μm) as calculated with (2).*

The condensation for a free droplet in an unlimited wet space occurs isotropically. This also applies if this droplet has an own charge [63]. As mentioned above, the condensation on droplets of a droplet cluster occurs anisotropically. From Fig. 4 it is evident that the presence of a strong external field introduces additional anisotropy into the condensation pattern. The velocity is minimal at the droplet's equator $\Theta = 0$ ($v_\perp = 0.009$ cm s$^{-1}$). Electro-condensation is most intense at the poles of the droplet, and it is maximally intensified at the negative pole (towards the electric field lines), because the droplet's own charge is negative. For the red line in Fig. 4, $v_\perp = 1.23$ cm s$^{-1}$ at the droplet's top and $v_\perp = 1.21$ cm s$^{-1}$ at its bottom. For the representation of a positively charged droplet, the figure should be turned upside down. A similar hypothesis about the asymmetry of mass exchange with the environment on a Leidenfrost droplet in an electric field was formulated by [64]. Simulation in [7] shows that the greatest increase in condensation growth is expected to occur when the electric field is oriented perpendicular to a flat water surface, which is similar to the conditions at the poles of our droplet. Molecular-dynamics simulations [65–66] even show that the anisotropy of the electric fields may lead to a stretching of the droplets along the field. We presume that the difference of our model and experimental data in upward ($U^-$) and downward ($U^+$) oriented fields (Fig. 3) are associated with the overlap of electro-condensation and the inhomogeneity of evaporation and condensation on the droplet surface.

## 5. Conclusions

Cloud droplets are hydrometeors that play key roles in processes associated with atmospheric electricity [67–70]. Numerous processes are not yet understood well enough for detailed incorporation into atmospheric physics models. This study focuses on the condensational growth of droplets exposed to external electrical fields. Both laboratory studies and a simplified physical model were applied. Laboratory experiments employed levitated, self-organized droplet clusters to study the growth of droplets in near-natural conditions. The external electric field enhanced condensational growth of levitated droplets by factors of up to two in a field of $1.5 \cdot 10^5$ V m$^{-1}$ strength. Growth rates could be measured precisely and their enhancement is in accordance with earlier studies [1–3, 5–8].

The condensation is accelerated in the electrical field because gaseous water molecules polarize and are accelerated. This is an electrically driven drift motion. As a result, condensation consists of a diffusion term and a drift term. The model that accounts for the droplet polarization is in qualitative agreement with the experimental data. The cause of quantitative differences presumably lies in an incomplete description of the steam-air flow, which supports the levitation of the droplet cluster. The experimentally observed difference in the growth rate of a droplet during the polarity reversal of the external field can be associated with the anisotropy of the phase transition processes on the droplet. This primary anisotropy is asymmetrical and not included in the physical model applied. The electrically driven condensation, that occurs almost symmetrically and non-uniformly (Fig. 4), stacks up with the primary one in the experiments. As a result, the physical model predicted growth rates that were by factors of 1.5 to 2.5 larger than the measured ones.

Further research is needed to find quantitative agreement between model and experimental data. We expect to develop further understanding of droplet growth mechanisms in natural clouds and fog. For example, the observation of bidirectional fluxes in fog lead to new questions of temporal limitations of droplet growth and evaporation mechanisms [71–72]. It is not likely that these bidirectional fluxes of microdroplets can be supported by electric fields that are typically present near the Earth's surface [11, 70].

**Acknowledgements**

The study was supported by the Russian Foundation for Basic Research (project No. 18-38-00232 mol_a), the Ministry of Education and Science of the Russian Federation (project No. 3.8191.2017/БЧ), and joint funding of the Ministry of Education and Science of the Russian Federation (project No. 3.12801.2018/12.2) and the German Academic Exchange Service (the Michael Lomonosov Programm – Linie A, 2018, No. 57391663). Reviewer comments on an earlier version of the manuscript are gratefully acknowledged.

**Appendix**
**Numerical calculation of electric fields**

The metal parts of the laboratory facility are quite distant, so that induced image charges, distorting the electrodes' field, can be entirely neglected. Therefore, the calculation of the fields is carried out in the geometry given in Fig. 1. In the current study, the bottom electrode is always grounded. A positive or negative potential is applied to the upper electrode.

The positive space potential generated by the upper electrode is defined as

$$\varphi^+(r,z) = \frac{1}{4\pi\varepsilon_0} \int_{z_1}^{z_2} \frac{\lambda^+ \, dl}{\sqrt{r^2 + (l-z)^2}}, \tag{A.1}$$

where $\varepsilon_0$ is the electric constant, $z_1 = 10.3 \, mm$ and $z_2 = 10.2 \, mm$ are coordinates describing the electrode position. Assuming that the charge distribution $\lambda^+$ over the surface of the electrode is quasi-uniform and thus constant, it can be taken out of the integral (A.1). The average surface charge density is thus computed employing the potential on the electrode surface $\varphi_0$:

$$\lambda^+ = \varphi_0 \left[ \frac{1}{4\pi\varepsilon_0} \ln \frac{r_i}{\sqrt{r_i^2 + (z_2-z_1)^2} - (z_2-z_1)} \right]^{-1},$$

where $r_i = 1.4 \, mm$ — the electrode radius.

A negative charge flows to the bottom grounded electrode:

$$\varphi_{-0}(r,z) = -\varphi^+(r,z).$$

The potential created by the bottom electrode is determined by

$$\varphi^-(r,z) = \frac{1}{4\pi\varepsilon_0} \int_{r_1}^{r_2} \frac{\lambda^- \, dl}{\sqrt{z^2 + (l-r)^2}}.$$

The average surface charge density is calculated from the potential $\varphi_{-0}$ on the electrode surface:

$$\lambda^+ = \varphi_{-0} \left[ \frac{1}{4\pi\varepsilon_0} \ln \frac{z_i}{\sqrt{z_i^2 + (r_2-r_1)^2} - (r_2-r_1)} \right]^{-1},$$

where $z_i = 0.7 \, mm$ — the thickness of the bottom electrode (washer), $r_1 = 2.25 \, mm$ and $r_2 = 4.45 \, mm$ are inner and outer radii of the washer.

The resulting scalar field and the field strength are calculated as follows

$$\varphi(r,z) = \varphi^+(r,z) + \varphi^-(r,z).$$

$$\vec{E} = -\nabla\varphi. \tag{A.2}$$

In the case of polarity reversal, the field calculation algorithm is similar. The calculation shows that in the space in which the cluster levitates the electric field can be considered as homogeneous, the strength drop across the diameter of one droplet does not exceed 0.02 V m$^{-1}$. This homogeneity is achieved by the large surface area of the upper electrode. Therefore, for analytical calculations, the field can be considered homogeneous. The calculation (A.2) showed that in our experiment (section 2.1) the electric field $E$ in the region of the droplet cluster is related to the potential $U$ on the upper electrode with the linear formula $E = U \cdot 218.8$ m$^{-1}$.

**Authors Contributions**

D.N.G. contributed to the further development of the experimental setup, developed the kinetic model, and drafted the manuscript. A.A.F. developed and carried out the experiment. N.E.A. calculated the electric fields strengths in the region of the levitation of the droplet cluster. O.K. reviewed and edited the manuscript. S.N.A. simulated electric fields near the surface of water droplet and verified analytical formulas for electric fields. All authors participated in the discussion of results and contributed to the manuscript.